\documentclass[preprint2,times,tighten]{aastex6}
\usepackage{amsmath,amstext}
\usepackage[all]{hypcap} 

\shorttitle{100 Kpc Distant Sgr Spur and Outer Virgo Overdensity}
\shortauthors{Sesar et al.}

\begin{document}

\title{The $>100$~kpc Distant Spur of the Sagittarius Stream and the Outer Virgo Overdensity, as Seen in PS1 RR Lyrae Stars}

\author{Branimir Sesar\altaffilmark{1}}
\author{Nina Hernitschek\altaffilmark{2}}
\author{Marion I.~P.~Dierickx\altaffilmark{3}}
\author{Mark A.~Fardal\altaffilmark{4}}
\author{Hans-Walter Rix\altaffilmark{1}}
\email{bsesar@mpia.de}

\altaffiltext{1}{Max Planck Institute for Astronomy, K\"{o}nigstuhl 17, D-69117 Heidelberg, Germany;}
\altaffiltext{2}{Division of Physics, Mathematics and Astronomy, Caltech, Pasadena, CA 91125}
\altaffiltext{3}{Astronomy Department, Harvard University, 60 Garden Street, Cambridge, MA 02138, USA}
\altaffiltext{4}{Space Telescope Science Institute, 3700 San Martin Drive, Baltimore, MD 21218, USA}

\begin{abstract}
We report the detection of spatially distinct stellar density features near the apocenters of the Sagittarius (Sgr) stream's main leading and trailing arm. These features are clearly visible in a high-fidelity stellar halo map that is based on RR Lyrae from Pan-STARRS1: there is a plume of stars 10~kpc beyond the apocenter of the leading arm, and there is a ``spur" extending to 130~kpc, almost 30~kpc beyond the previously detected apocenter of the trailing arm. Such apocenter substructure is qualitatively expected in any Sgr stream model, as stars stripped from the progenitor at different pericenter passages become spatially separated there. The morphology of these new Sgr stream substructures could provide much-needed new clues and constraints for modeling the Sgr system, including the level of dynamical friction that Sgr has experienced. We also report the discovery of a new, presumably unrelated halo substructure at 80 kpc from the Sun and $10^\circ$ from the Sgr orbital plane, which we dub the Outer Virgo Overdensity.
\end{abstract}

\maketitle

\section{Introduction}

The Sagittarius stellar stream (Sgr; \citealt{iba01}) is the 800-pound gorilla of stellar substructure in the Galactic halo, containing at least an order of magnitude more stars than other distinct stellar streams (e.g., Orphan or Cetus Polar stream; \citealt{gri06,new09}). As the progenitor Sgr galaxy was a rather massive, luminous satellite galaxy ($L\approx10^8$ $L_\sun$, \citealt{no10}; $M>10^9$ $M_\sun$, \citealt{pen10}), the stellar stream resulting from its tidal disruption has spread quickly, and therefore widely, in stream angle near the orbit of the progenitor: it is the only known stream that wraps more than once around the Galaxy \citep{maj03}. Sgr is on a high-latitude orbit, spanning galactocentric distances from $\approx15$~kpc to $\approx 100$~kpc along its orbit \citep{bel14}. These properties should make Sgr an excellent system for constraining the gravitational potential of the Galactic halo, its radial profile and shape (e.g., its flattening and degree of triaxiality). And even though dynamical models for the Sgr stream have been constructed for over a decade \citep{hel04,lm10,gbe14,dl17}, unambiguous and detailed inferences about the Galactic halo potential beyond an estimate of the Milky Way mass enclosed within the orbit of the Sgr stream have been difficult to obtain.

Detailed modeling of the Sgr stream has been challenging, because its position-velocity distribution depends on three aspects: the Galactic potential, the progenitor orbit, and the internal structure of the progenitor \citep{pen10,gbe14}.  Stellar tidal streams form their leading and trailing arms when progenitor stars become unbound, foremost as the consequence of the progenitor's pericenter passages (e.g., see Figure 3 of \citealt{bov14}). But at any point along the arms there is a mix of stream stars that have become unbound on different orbits and at different points in the past, often different pericenter passages (e.g., see bottom right panel of Figure 5 of \citealt{bov14}). This is presumably manifested in the well-established ``bifurcation" \citep{bel06,kop12} of the Sgr stream. At apocenter, where the velocities are the lowest and hence most similar among stream stars, these orbit differences among stream stars manifest themselves mostly in spatial (not velocity) differences. As a consequence, the Sgr stream is expected to show distinct spatial structure near the apocenters of the leading and trailing arms (e.g., top left panel of Figure 10 of \citealt{far15}). But the Sgr stream structure at the apocenters has not yet been mapped out in detail.

Here we show that the high-fidelity map of the Sgr stream, enabled by the recent, extensive set of RR Lyrae stars from Pan-STARRS1 \citep{ses17}, clearly reveals such stream structures. These observed substructures are in a good qualitative agreement with recent dynamical models \citep{gbe14,far15,dl17}, and can serve as a qualitatively new constraint on Sgr stream models. The bifurcation of the Sgr stream, which is another well-known feature of the stream \citep{bel06,kop12,sla13}, is discussed in a companion paper (Hernitschek, N.~et al., in prep.), along with a more quantitative description of the Sgr tidal stream.

The remainder of this paper is organized as follows: \autoref{sec:data} briefly recapitulates the data set underlying the results; \autoref{sec:results} shows the observational evidence of spatial substructure near the apocenters of both arms in the Sgr stream; \autoref{sec:VOD} describes a newly discovered halo overdensity; \autoref{sec:modelcomparison} discusses the qualitative agreement of observed substructures with recent models, and \autoref{sec:conclusions} summarizes our conclusions.

\section{PS1 RR Lyrae Stars}\label{sec:data}

Our analysis uses a sample of highly probable type $ab$ RR Lyrae stars (hereafter RRab), selected from the \citet{ses17} catalog of RR Lyrae stars. According to \citet[see their Sections~5 and~6]{ses17}, 90\% of objects in this sample are expected to be true RRab stars (i.e., the purity of the sample is 90\%). At high galactic latitudes ($|b| > 10\arcdeg$) the sample is expected to contain at least 80\% of all RRab stars to 80 kpc from the Sun (i.e., the completeness of the sample is at least 80\%; see their Figure~11 for the sample completeness as a function of the $r$-band magnitude). The distance modulus uncertainties are $\sigma_{\rm DM} = 0.06(rnd) \pm0.03(sys)$ mag, corresponding to a distance precision of $\sim3\%$, as measured by \citet[see their Section~3.3]{ses17}.

The Pan-STARRS1 (PS1) catalog of RR Lyrae stars covers about three quarters of the sky (i.e., the sky north of declination $-30\arcdeg$). But since the focus of the current work is the Sgr tidal stream, we limit the sample to $\approx19,000$ RRab stars located within $13\arcdeg$ of the Sgr orbital plane (i.e., $|\tilde{B}_{\sun}|<13\arcdeg$), where $\tilde{\Lambda}_{\sun}$ and $\tilde{B}_{\sun}$ are heliocentric Sagittarius coordinates as defined by \citet{bel14}. In this coordinate system, the equator $\tilde{B}_{\sun} = 0\arcdeg$ is aligned with the plane of the stream.

\section{Apocenter Substructure in the Sgr Stream}\label{sec:results}

\begin{figure*}
\epsscale{0.7}
\plotone{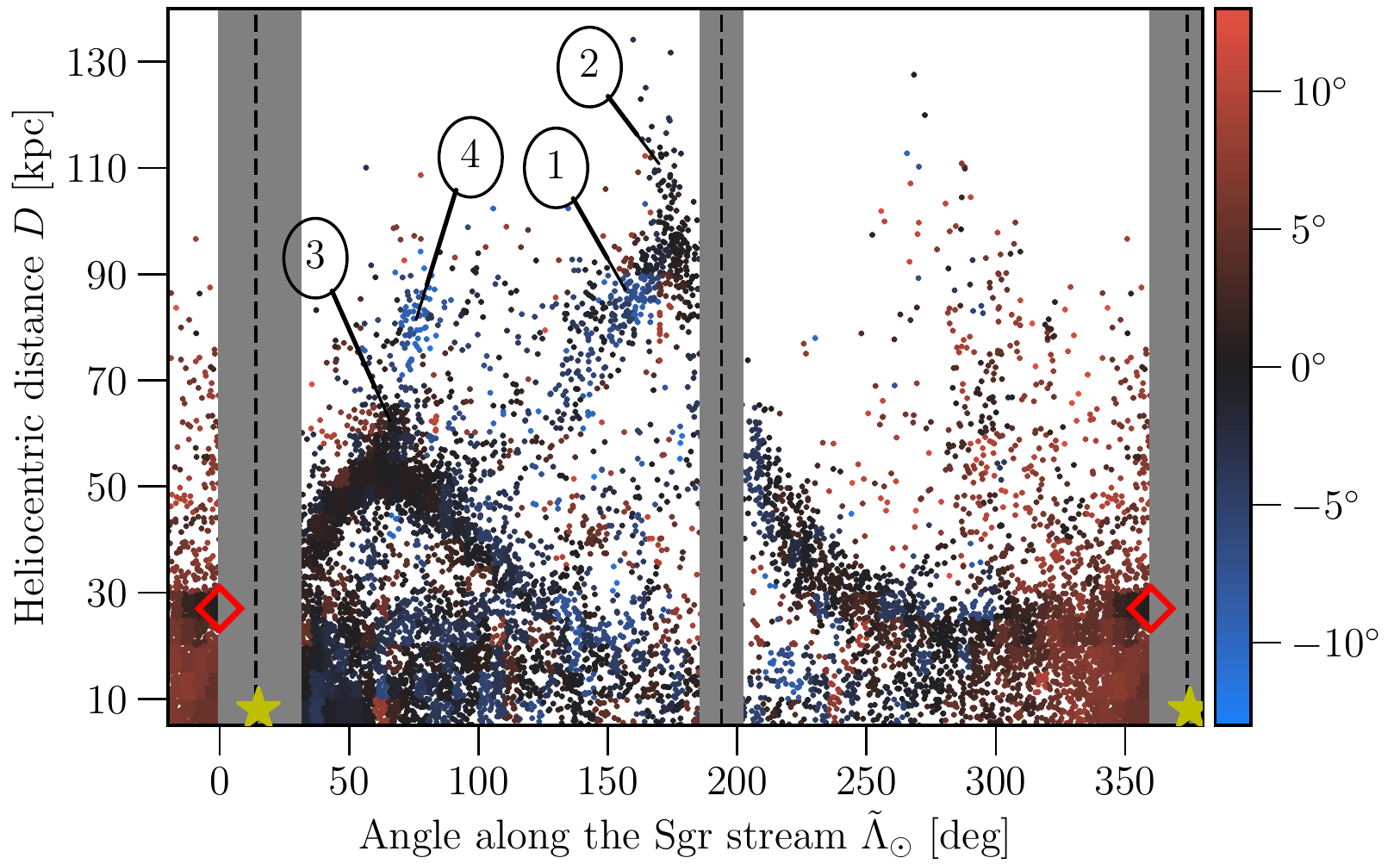}

\plotone{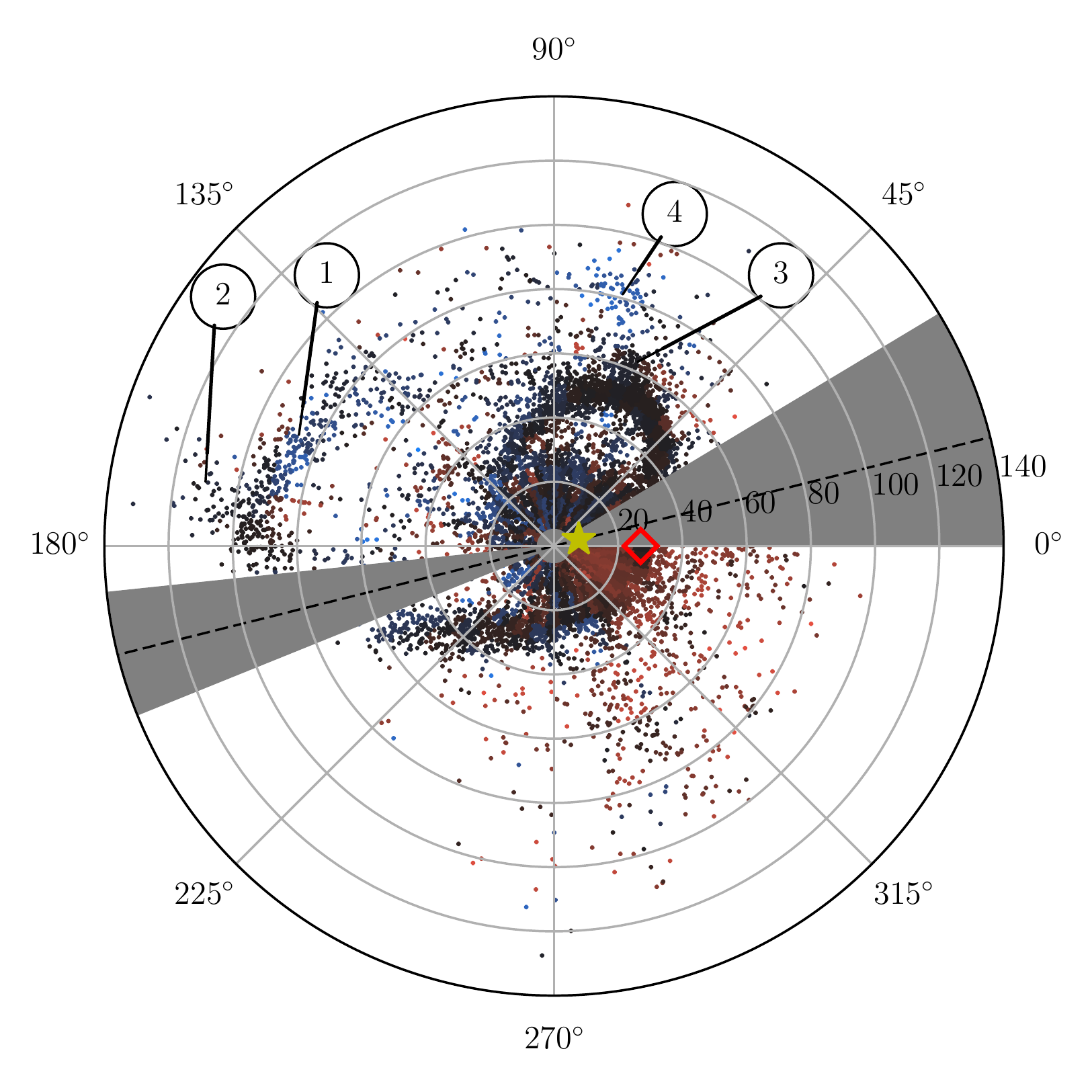}
\caption{
Distribution of PS1 RRab stars within $13\arcdeg$ of the Sgr orbital plane, shown in two different projections. The color indicates the stars' angular distance from the Sgr orbital plane (in $\tilde{\Lambda}_{\sun}$ and $D$ coordinates). The gray areas indicate regions near the Galactic plane (dashed lines) where the PS1 RR Lyrae sample is significantly incomplete. The positions of the Galactic center and the Sgr dSph galaxy are shown by the yellow star and red open diamond symbols, respectively. The Sun is located at $(\tilde{\Lambda}_{\sun}, D)=0\arcdeg, 0$ kpc. The features annotated by numbers are discussed in \autoref{sec:results}.
\label{Sgr_cartesian_obs_only}}
\end{figure*}

\begin{figure*}
\epsscale{0.8}
\plotone{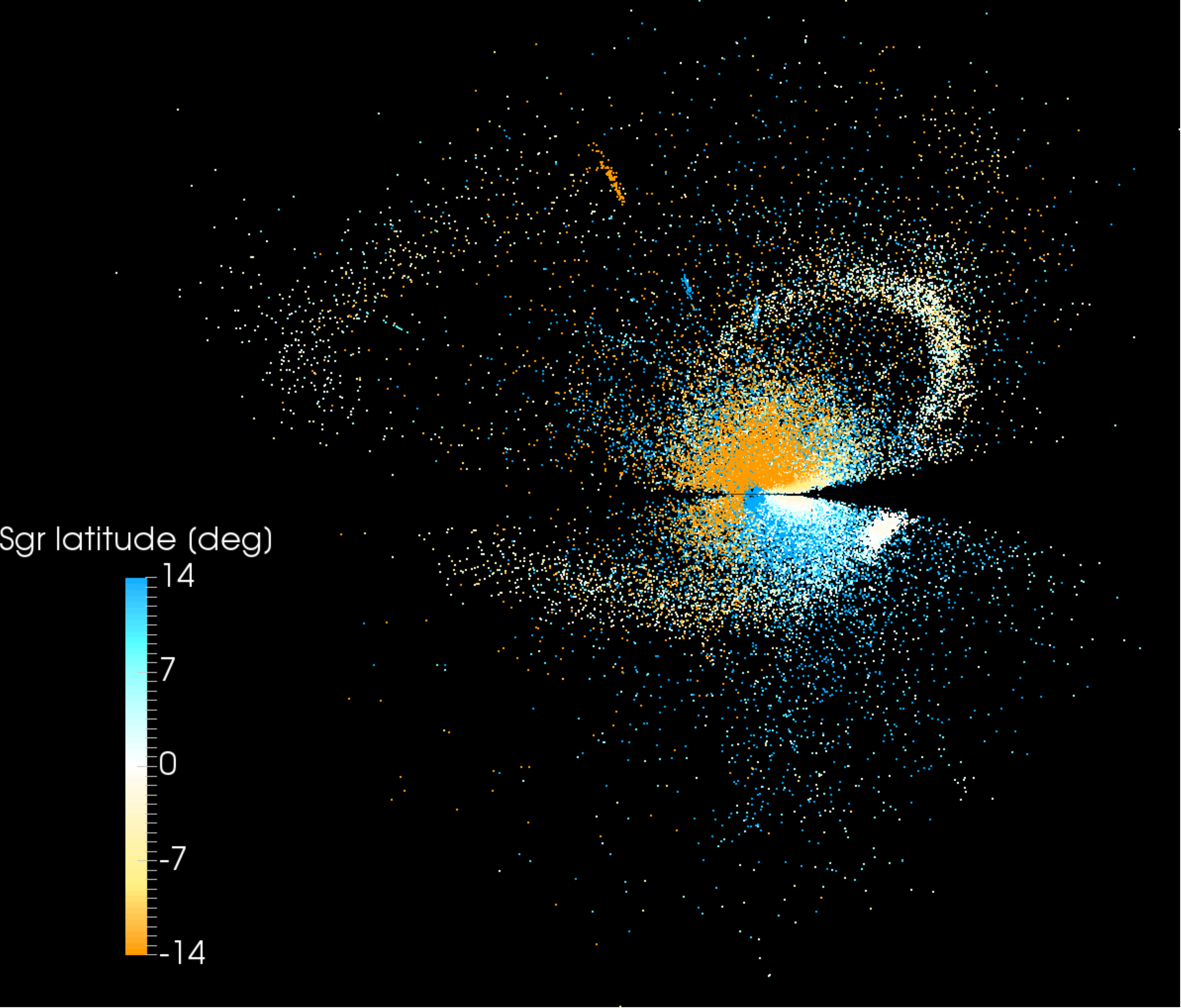}
\caption{
Snapshot of an animation that shows the 3D distribution of $\approx44,000$ highly probable RRab stars selected by \citet{ses17} from the PS1 $3\pi$ survey. The color indicates the angular distance of a star from the Sgr orbital plane (i.e., its Sgr latitude $\tilde{B}_{\sun}$). The extended overdensities are dSph galaxies (e.g., Draco, Sextans, Ursa Minor). The animation is provided in the electronic version of the Journal.
\label{animation_snapshot}}
\end{figure*}

The distribution of PS1 RRab stars within $13\arcdeg$ of the Sgr orbital plane is shown in \autoref{Sgr_cartesian_obs_only}, and the distribution of {\it all} PS1 RRab is illustrated in \autoref{animation_snapshot}. The wide and deep coverage by the PS1 $3\pi$ survey \citep{cha11}, and the purity and completeness of the PS1 RRab sample, now provide an almost complete, $360\arcdeg$ view of the Sgr stream (in RRab) from 5 kpc to at least 120 kpc from the Sun, with only a minor loss of coverage (due to dust extinction) in regions close to the Galactic plane.

For the first time, these data clearly show the trailing arm as it extends (right to left in the top panel of \autoref{Sgr_cartesian_obs_only}) from $D\approx25$ kpc at $\tilde{\Lambda}_{\sun}=0\arcdeg$ (i.e., the core of the Sgr dSph) to well beyond its apocenter at $D\approx90$ kpc and $\tilde{\Lambda}_{\sun}\approx170\arcdeg$. At the trailing arm's apocenter the stream appears to fork into two branches: one turning back towards the Galactic center (feature 1 in \autoref{Sgr_cartesian_obs_only}), and the other extending as far as 120 kpc from the Sun (feature 2 in \autoref{Sgr_cartesian_obs_only}).

Initially detected by \citet{new03}, the part of the trailing arm that is turning back towards the Galactic center (feature 1) was also identified by \citet{dra13}, but deemed a new stellar stream, those authors named Gemini. However, our wide view clearly shows that feature 1 is not a new stream, but simply a part of the Sgr stream as proposed by \citet{new03} and \citet{bel14}.

Our data also allow us to identify the spur extending to $\ga120$ kpc of the Sun (feature 2 in \autoref{Sgr_cartesian_obs_only}) as a new branch of the Sgr stream. \citet{dra13} have detected four RRab stars in this region (at $(\tilde{\Lambda}_{\sun}, D)=172\arcdeg, 120$ kpc, see their Figure~7), but did not recognize it as a new part of the Sgr stream.

Near the position of the leading arm's apocenter ($\tilde{\Lambda}_{\sun}\approx60\arcdeg$), but about 10 kpc further, there is an overdensity of RRab stars that we have labeled as ``feature 3'' in \autoref{Sgr_cartesian_obs_only}. While it appears distinct in the $(D,\tilde{\Lambda}_{\sun})$-plane, this group of stars is located almost exactly in the Sgr orbital plane (i.e, $\tilde{B}_{\sun}\approx0\arcdeg$), as the color coding in \autoref{Sgr_cartesian_obs_only} shows. Due to its proximity to the Sgr orbital plane, its position relative to the apocenter of the leading arm, and the qualitative existence of such an apocenter lump in most Sgr simulations (see \autoref{sec:modelcomparison}), we believe that this feature is associated with the Sgr stream. Such spurs and lumps are a generic model prediction for the disruption of the Sgr dSph galaxy, and we discuss it (and other features) in more detail in \autoref{sec:modelcomparison}.

\section{Outer Virgo Overdensity}\label{sec:VOD}

Another notable feature is a clump of stars at $(\tilde{\Lambda}_{\sun}, D)=75\arcdeg, 80$ kpc (feature 4 \autoref{Sgr_cartesian_obs_only}). This clump is clearly offset from the plume near the leading arm apocenter (i.e., feature 3), by about 15 kpc in the radial direction and by about $9\arcdeg$ from the Sgr orbital plane (or $\approx13$~kpc). As no simulation predicts Sgr debris at that position (see \autoref{obs_sim_comparison}), we believe that this clump of RRab stars traces a new halo substructure, which we name the Outer Virgo Overdensity (Outer VOD; based on its location in the Virgo constellation and to distinguish it from the (inner) Virgo Overdensity at $\approx6-20$ kpc of the Sun; \citealt{viv01,new02,jur08}). Obviously, kinematics of these stars will help settle this question.

\begin{figure*}
\epsscale{0.8}
\plotone{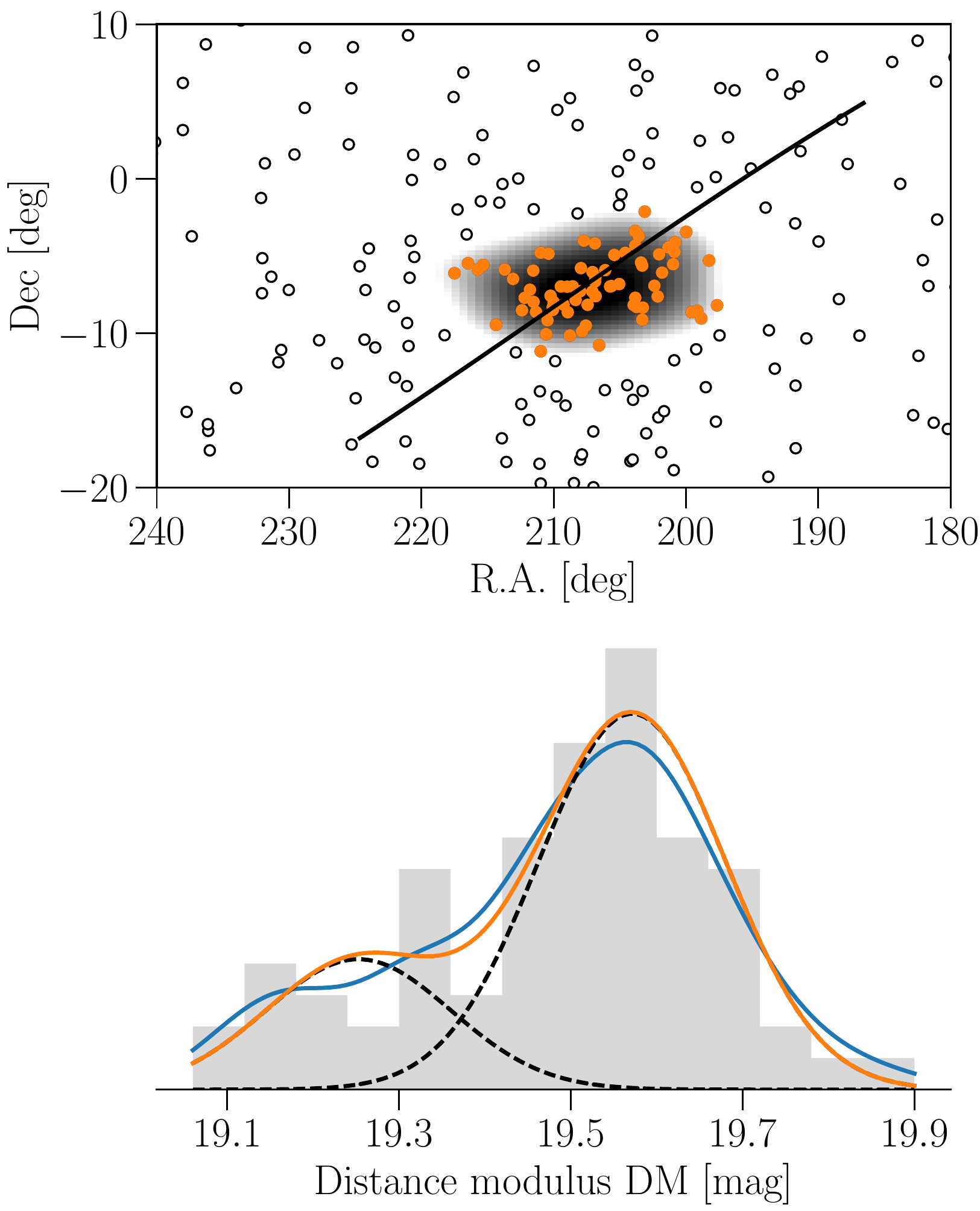}
\caption{
{\em Top}: Spatial distribution of PS1 RRab stars (circles) in the vicinity of the newly discovered Outer Virgo Overdensity (VOD). The gray patch indicates the location and the extent of the overdensity, and the orange solid circles show the positions of RRab stars that are spatially consistent with the overdensity. The $\tilde{B}_{\sun}=-9\arcdeg$ latitude in the Sgr coordinate system is illustrated by the solid line. {\em Bottom}: Distribution of Outer VOD RRab stars (orange solid circles in the top panel) in distance modulus, as measured by 1) binning the data in 0.06 mag bins (gray histogram), 2) running Gaussian kernel density estimation on the data (blue solid line, kernel bandwidth is 0.06 mag), and 3) by fitting two Gaussians to the data (dashed lines, the orange solid line shows their sum). The optimal kernel bandwidth and the number of Gaussians were determined using 10-fold cross-validation.
\label{Outer_Virgo_radec}}
\end{figure*}

To examine the Outer VOD in more detail, in \autoref{Outer_Virgo_radec} we show the distribution of PS1 RRab stars near $\tilde{\Lambda}_{\sun}\approx75\arcdeg$ and between 65 and 95 kpc of the Sun. While a clump of stars is easily visible near ${\rm (R.A., Dec) = }207\arcdeg, -7\arcdeg$, the full extent of the overdensity is more difficult to discern. To trace the overdensity, we measure the density of RRab stars implied by their spatial distribution by using a bivariate Gaussian kernel\footnote{The optimal bandwidth of the kernel was determined by using leave-one-out cross-validation.}, and then plot the measured density as a gray scale patch in \autoref{Outer_Virgo_radec} (the gray scale has been tweaked to emphasize the location and the extent of the Outer VOD). The overdensity covers $\approx150$ deg$^2$ and about 70 RRab stars are spatially consistent with it. Their position and other relevant information are provided in \autoref{table1}.

\capstartfalse
\begin{deluxetable}{ccccccc}
\tabletypesize{\scriptsize}
\tablecolumns{7}
\tablecaption{Outer Virgo Overdensity PS1 RR Lyrae Stars\label{table1}}
\tablehead{
\colhead{R.A.} & \colhead{Decl.} & \colhead{$score_{\rm 3,ab}^a$} & \colhead{DM$^b$} & \colhead{Period} & \colhead{$\phi_0^c$} & \colhead{$A_r^d$} \\
\colhead{(deg)} & \colhead{(deg)} & \colhead{$ $} & \colhead{(mag)} & \colhead{(day)} & \colhead{(day)} & \colhead{(mag)}
}
\startdata
197.65117 & -8.20577 & 0.85 & 19.44 & 0.522410 &  0.32399 & 0.97 \\
198.84621 & -9.05291 & 0.82 & 19.70 & 0.745662 &  0.26986 & 0.43 \\
199.15347 & -8.56380 & 0.99 & 19.32 & 0.612302 & -0.23730 & 0.65
\enddata
\tablenotetext{a}{Final RRab classification score.}
\tablenotetext{b}{Distance modulus. The uncertainty in distance modulus is $0.06(rnd)\pm0.03(sys)$ mag.}
\tablenotetext{c}{Phase offset (see Equation 2 of \citealt{ses17}).}
\tablenotetext{d}{PS1 $r$-band light curve amplitude.}
\tablecomments{A machine readable version of this table is available in the electronic edition of the Journal. A portion is shown here for guidance regarding its form and content.} 
\end{deluxetable}
\capstarttrue

As the bottom panel of \autoref{Outer_Virgo_radec} shows, the distribution of Outer VOD RRab stars in distance modulus can be modeled as a sum of two Gaussians centered at 19.57 mag and 19.25 mag, with amplitudes of 0.74 and 0.26, respectively. The standard deviation of these Gaussians is 0.11 mag, implying an intrinsic scatter or line-of-sight depth of 0.09 mag (once 0.06 mag of uncertainty in DM is subtracted in quadrature). Thus, the Outer VOD seems have a line-of-sight size ($1\sigma$) of $\approx 4$~kpc, comparable to its extent projected onto the sky. We use a sum of Gaussians only as a simple model to describe the distribution of Outer VOD RR Lyrae stars along the line of sight, and do not attach any physical interpretation to the Gaussians at this point.

\section{Qualitative Comparison to Models}\label{sec:modelcomparison}

In \autoref{sec:results}, we have used the position and morphology of features 2 and 3 to tentatively associate them with the Sgr tidal stream. The $N$-body models of \citet{gbe14}, \citet{far15}, and \citet{dl17}, which describe the disruption of the Sgr dwarf spheroidal galaxy, now provide us with an opportunity to explore the extent to which similar features also exist in simulations.

\begin{figure*}
\epsscale{0.8}
\plotone{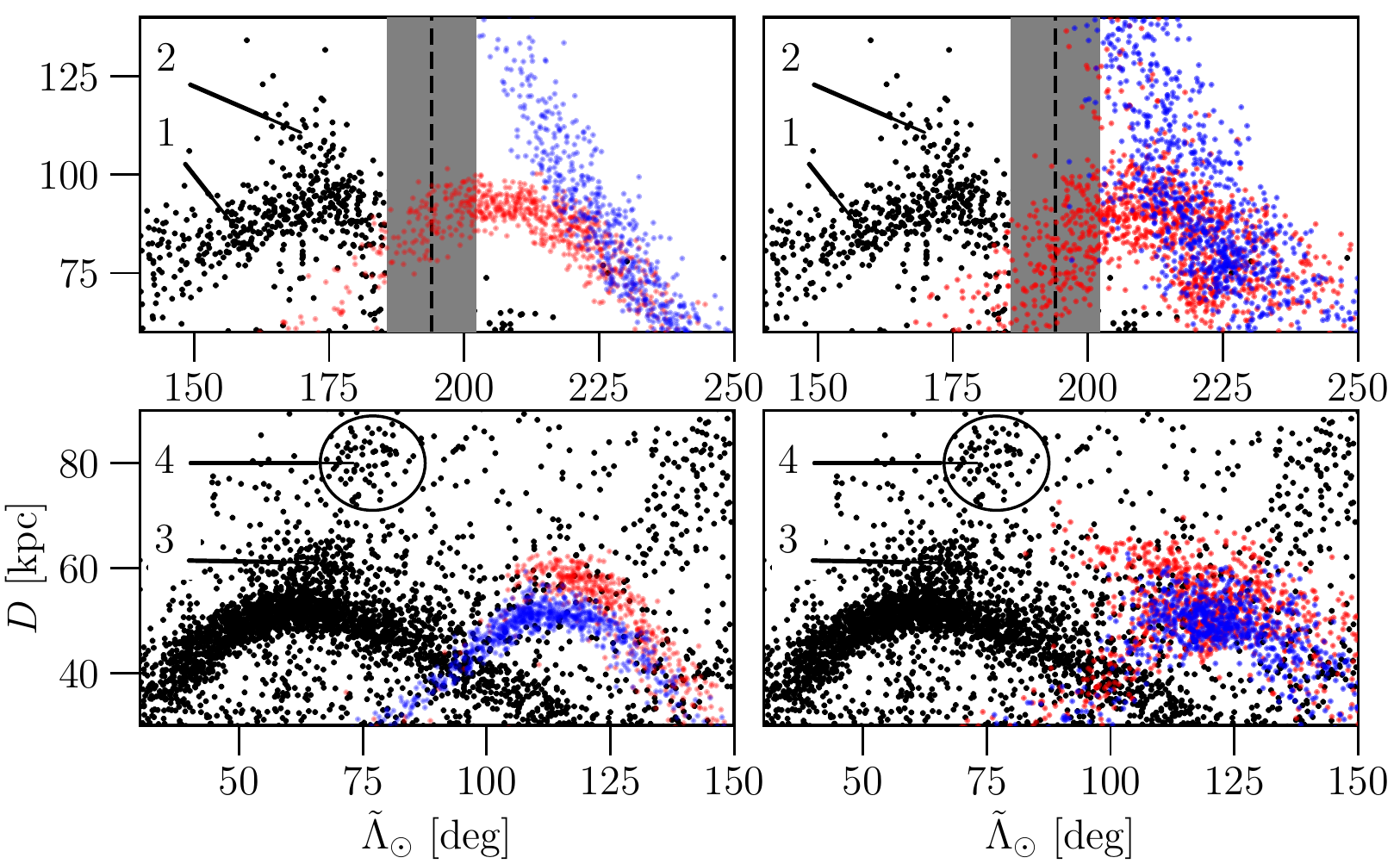}
\caption{
Side-by-side comparison of the observed (black points) and simulated Sgr streams (blue and red points; left panels, \citealt{far15}; right panels, \citealt{dl17}). The simulated streams have been offset from their original positions to enable easier comparison (by $\Delta\tilde{\Lambda}_{\sun}=40\arcdeg$ and $\Delta\tilde{\Lambda}_{\sun}=70\arcdeg$ in the top and bottom panels, respectively). The blue and red points show $N-$body particles stripped off the Sgr dSph galaxy during pericenter passages that happened about 1.3 and 2.7 Gyr ago, respectively. To mimic the uncertainty in distances to PS1 RRab stars, we have added $0.03\times D$ of Gaussian noise to distances of $N-$body particles.
\label{obs_sim_comparison}}
\end{figure*}

\autoref{obs_sim_comparison} compares features 2 and 3 observed with PS1 RRab stars, with similar features observed in $N$-body simulations of \citet{far15} and \citet{dl17}. We find that the $>100$ kpc spur (top panels) and the plume of stars near the apocenter of the leading arm (bottom panels) have similar counterparts in $N$-body simulations. The same features are also present in the \citet{gbe14} simulation (see their Figure 6).

Qualitatively, the observed spur seems to match its simulated counterparts in position and width. The rising incompleteness of the PS1 RRab sample at distances greater than 80 kpc makes a more quantitative comparison difficult. Feature 3,  the observed plume of stars near the leading arm apocenter appears to be narrower and more distinct from the rest of the stream than in \citet{far15} or \citet{dl17} simulations (along the $\tilde{\Lambda}_{\sun}$ direction). This difference may indicate a mismatch between the Galactic potential and the potential assumed in simulations, since the breadth of the apocentric region is strongly influenced by the spread of angular momentum at a given energy (e.g., see Figure 1 of \citealt{jsb01}), which in turn is strongly affected by the shape of the potential.

This comparison also shows that the Outer Virgo Overdensity (i.e., feature 4 in \autoref{Sgr_cartesian_obs_only} and circled in the bottom panel of \autoref{obs_sim_comparison}) has no counterpart in either simulation, suggesting that it is not tidal debris associated with the Sgr dSph galaxy.

\section{Discussion and Conclusions}\label{sec:conclusions}

We have presented what is currently the deepest, widest, and most precise map of the Sagittarius (Sgr) tidal stream, traced using a sample of $\approx19,000$ Pan-STARRS1 RRab stars. Thanks to the high purity and completeness of this sample (90\% and 80\%, respectively), and distances that are precise at the 3\% level \citep{ses17}, we were able to identify two new spatial substructures in the stream:
\begin{itemize}
\item A spur extending from the Sgr trailing arm apocenter to at least 120 kpc of the Sun (feature 2 in \autoref{Sgr_cartesian_obs_only} and~\autoref{obs_sim_comparison}), and
\item A plume of stars near the leading arm apocenter (and offset by about 10 kpc from it; feature 3 in \autoref{Sgr_cartesian_obs_only} and~\autoref{obs_sim_comparison}).
\end{itemize}

Through a qualitative comparison (\autoref{obs_sim_comparison}) we have found that the above apocenter features are also present in $N$-body simulations of \citet{gbe14}, \citet{far15}, and \citet{dl17}, that model the disruption of the Sgr dSph. According to these simulations, the $>100$ kpc spur is the continuation of the trailing arm that formed during a pericenter passage that happened about 1.3 Gyr ago. The plume of stars near the leading arm apocenter (feature 3), on the other hand, is dynamically older and indicates the apocenter of the leading arm that formed during a pericenter passage about 2.7 Gyr ago.

These new Sgr stream substructures could provide much-needed new clues and constraints for modeling the Galactic potential and the dynamical evolution of the Sgr system. For example, due to the radial extension of the $>100$ kpc Sgr spur, the heliocentric distances and line of sight velocities of its stars can be used to directly measure the gravitational potential of the Milky Way at large radii ($>90$ kpc), where it is very poorly constrained (e.g., see Figure 13 of \citealt{kue15}). The outer plume of stars near the leading arm apocenter, presumably stripped during an earlier pericenter passage, is clearly separated in orbital energy from the dominant component at that apocenter. Dynamical friction will shift the energies and orbital periods of these two separate components, suggesting that we may constrain the orbital decay of the Sgr dwarf's orbit by carefully observing the spatial distribution of the two components.

While we have observed the Sgr stream to reach as far as 120 kpc from the Sun (feature 2), the stream may extend much further than that. For example, the simulations of \citet{dl17} predict that the most distant arm of the stream, which they label the ``northwest branch'', extends to distances $>200$ kpc (see their Figures 8 and 10). The fact that we even detect the Sgr stream up to 120 kpc, provides some support for this claim. At 120 kpc, the completeness of the \citet{ses17} RRab sample is expected to be $\lesssim10\%$ (see their Figure 11). If the Sgr stream was ending at 120 kpc, its surface density should already be fairly low by that point and given the expected completeness of the RRab sample, we likely would not be able to detect it at all.

How far does the Sgr spur (i.e., feature 2) extend? The identification of distant RR Lyrae stars in this region requires deep ($>23$ mag) multi-epoch imaging ($>30$ observations) that none of the planned wide-area optical surveys can provide. For example, the Zwicky Transient Facility (ZTF; \citealt{ztf}) is too shallow (median single-visit depth $r\approx20.4$ mag), and the Large Synoptic Survey Telescope (LSST; \citealt{ive08}), while deep, will not observe the sky north of declination $10\arcdeg$ (the observed part of feature 2 is located between $110\arcdeg < R.A. < 125\arcdeg$ and $25\arcdeg < Dec < 30\arcdeg$). Instead of a wide-area survey, a small survey with a wide-field imager on a large telescope (e.g., the Hyper Suprime-Cam on 8.2-m Subaru telescope; \citealt{miy12}), may be a much better choice.

In addition to detecting two substructures associated with the Sgr tidal stream, we have also detected a new halo overdensity at a distance of 80 kpc, which we have named the Outer Virgo Overdensity (Outer VOD). Due to its clear separation from the Sgr stream ($\approx20$ kpc in the radial direction and $10\arcdeg$ of the Sgr orbital plane), we do not believe it is associated with the Sgr stream. Instead, we speculate that the Outer VOD overdensity may be a remnant of a disrupted dSph galaxy or a globular cluster that was accreted independently by the Milky Way or perhaps together with the Sgr dSph galaxy (i.e., it may have been a satellite of the Sgr dSph). A spectroscopic study of RRab stars associated with this substructure (see \autoref{table1}) would be very useful, as the radial velocities and metallicities would constrain its nature and orbital parameters (i.e., dSph vs. globular cluster, satellite of Sgr or not).

\acknowledgments

B.S.~would like to dedicate this work to his son, Elon. B.S.~would also like to thank his family, friends, and colleagues for supporting him over 15 years in Astronomy. B.S, N.H.~and H.-W.R.~acknowledge funding from the European Research Council under the European Union’s Seventh Framework Programme (FP 7) ERC Grant Agreement n.~${\rm [321035]}$. M.F.~acknowledges support through HST grant GO-13443. The Pan-STARRS1 Surveys (PS1) have been made possible through contributions by the Institute for Astronomy, the University of Hawaii, the Pan-STARRS Project Office, the Max-Planck Society and its participating institutes, the Max Planck Institute for Astronomy, Heidelberg and the Max Planck Institute for Extraterrestrial Physics, Garching, The Johns Hopkins University, Durham University, the University of Edinburgh, the Queen's University Belfast, the Harvard-Smithsonian Center for Astrophysics, the Las Cumbres Observatory Global Telescope Network Incorporated, the National Central University of Taiwan, the Space Telescope Science Institute, and the National Aeronautics and Space Administration under Grant No.~NNX08AR22G issued through the Planetary Science Division of the NASA Science Mission Directorate, the National Science Foundation Grant No.~AST-1238877, the University of Maryland, Eotvos Lorand University (ELTE), and the Los Alamos National Laboratory. 

\bibliography{main}

\begin{thebibliography}{}
\expandafter\ifx\csname natexlab\endcsname\relax\def\natexlab#1{#1}\fi

\bibitem[{{Bellm}(2014)}]{ztf}
{Bellm}, E. 2014, in The Third Hot-wiring the Transient Universe Workshop, ed.
  P.~R. {Wozniak}, M.~J. {Graham}, A.~A. {Mahabal}, \& R.~{Seaman}, 27--33

\bibitem[{{Belokurov} {et~al.}(2006){Belokurov}, {Zucker}, {Evans}, {Gilmore},
  {Vidrih}, {Bramich}, {Newberg}, {Wyse}, {Irwin}, {Fellhauer}, {Hewett},
  {Walton}, {Wilkinson}, {Cole}, {Yanny}, {Rockosi}, {Beers}, {Bell},
  {Brinkmann}, {Ivezi{\'c}}, \& {Lupton}}]{bel06}
{Belokurov}, V., {Zucker}, D.~B., {Evans}, N.~W., {et~al.} 2006, \apjl, 642,
  L137

\bibitem[{{Belokurov} {et~al.}(2014){Belokurov}, {Koposov}, {Evans},
  {Pe{\~n}arrubia}, {Irwin}, {Smith}, {Lewis}, {Gieles}, {Wilkinson},
  {Gilmore}, {Olszewski}, \& {Niederste-Ostholt}}]{bel14}
{Belokurov}, V., {Koposov}, S.~E., {Evans}, N.~W., {et~al.} 2014, \mnras, 437,
  116

\bibitem[{{Bovy}(2014)}]{bov14}
{Bovy}, J. 2014, \apj, 795, 95

\bibitem[{{Chambers}(2011)}]{cha11}
{Chambers}, K.~C. 2011, in Bulletin of the American Astronomical Society,
  Vol.~43, American Astronomical Society Meeting Abstracts \#218, 113.01

\bibitem[{{Dierickx} \& {Loeb}(2017)}]{dl17}
{Dierickx}, M.~I.~P., \& {Loeb}, A. 2017, \apj, 836, 92

\bibitem[{{Drake} {et~al.}(2013){Drake}, {Catelan}, {Djorgovski}, {Torrealba},
  {Graham}, {Mahabal}, {Prieto}, {Donalek}, {Williams}, {Larson},
  {Christensen}, \& {Beshore}}]{dra13}
{Drake}, A.~J., {Catelan}, M., {Djorgovski}, S.~G., {et~al.} 2013, \apj, 765,
  154

\bibitem[{Fardal {et~al.}(2015)Fardal, Huang, \& Weinberg}]{far15}
Fardal, M.~A., Huang, S., \& Weinberg, M.~D. 2015, Monthly Notices of the Royal
  Astronomical Society, 452, 301

\bibitem[{{Gibbons} {et~al.}(2014){Gibbons}, {Belokurov}, \& {Evans}}]{gbe14}
{Gibbons}, S.~L.~J., {Belokurov}, V., \& {Evans}, N.~W. 2014, \mnras, 445, 3788

\bibitem[{{Grillmair}(2006)}]{gri06}
{Grillmair}, C.~J. 2006, \apjl, 645, L37

\bibitem[{{Helmi}(2004)}]{hel04}
{Helmi}, A. 2004, \apjl, 610, L97

\bibitem[{{Ibata} {et~al.}(2001){Ibata}, {Irwin}, {Lewis}, \& {Stolte}}]{iba01}
{Ibata}, R., {Irwin}, M., {Lewis}, G.~F., \& {Stolte}, A. 2001, \apjl, 547,
  L133

\bibitem[{{Ivezic} {et~al.}(2008){Ivezic}, {Tyson}, {Abel}, {Acosta},
  {Allsman}, {AlSayyad}, {Anderson}, {Andrew}, {Angel}, {Angeli}, {Ansari},
  {Antilogus}, {Arndt}, {Astier}, {Aubourg}, {Axelrod}, {Bard}, {Barr},
  {Barrau}, {Bartlett}, {Bauman}, {Beaumont}, {Becker}, {Becla}, {Beldica},
  {Bellavia}, {Blanc}, {Blandford}, {Bloom}, {Bogart}, {Borne}, {Bosch},
  {Boutigny}, {Brandt}, {Brown}, {Bullock}, {Burchat}, {Burke}, {Cagnoli},
  {Calabrese}, {Chandrasekharan}, {Chesley}, {Cheu}, {Chiang}, {Claver},
  {Connolly}, {Cook}, {Cooray}, {Covey}, {Cribbs}, {Cui}, {Cutri}, {Daubard},
  {Daues}, {Delgado}, {Digel}, {Doherty}, {Dubois}, {Dubois-Felsmann},
  {Durech}, {Eracleous}, {Ferguson}, {Frank}, {Freemon}, {Gangler}, {Gawiser},
  {Geary}, {Gee}, {Geha}, {Gibson}, {Gilmore}, {Glanzman}, {Goodenow},
  {Gressler}, {Gris}, {Guyonnet}, {Hascall}, {Haupt}, {Hernandez}, {Hogan},
  {Huang}, {Huffer}, {Innes}, {Jacoby}, {Jain}, {Jee}, {Jernigan},
  {Jevremovic}, {Johns}, {Jones}, {Juramy-Gilles}, {Juric}, {Kahn}, {Kalirai},
  {Kallivayalil}, {Kalmbach}, {Kantor}, {Kasliwal}, {Kessler}, {Kirkby},
  {Knox}, {Kotov}, {Krabbendam}, {Krughoff}, {Kubanek}, {Kuczewski},
  {Kulkarni}, {Lambert}, {Le Guillou}, {Levine}, {Liang}, {Lim}, {Lintott},
  {Lupton}, {Mahabal}, {Marshall}, {Marshall}, {May}, {McKercher}, {Migliore},
  {Miller}, {Mills}, {Monet}, {Moniez}, {Neill}, {Nief}, {Nomerotski},
  {Nordby}, {O'Connor}, {Oliver}, {Olivier}, {Olsen}, {Ortiz}, {Owen}, {Pain},
  {Peterson}, {Petry}, {Pierfederici}, {Pietrowicz}, {Pike}, {Pinto}, {Plante},
  {Plate}, {Price}, {Prouza}, {Radeka}, {Rajagopal}, {Rasmussen}, {Regnault},
  {Ridgway}, {Ritz}, {Rosing}, {Roucelle}, {Rumore}, {Russo}, {Saha},
  {Sassolas}, {Schalk}, {Schindler}, {Schneider}, {Schumacher}, {Sebag},
  {Sembroski}, {Seppala}, {Shipsey}, {Silvestri}, {Smith}, {Smith}, {Strauss},
  {Stubbs}, {Sweeney}, {Szalay}, {Takacs}, {Thaler}, {Van Berg}, {Vanden Berk},
  {Vetter}, {Virieux}, {Xin}, {Walkowicz}, {Walter}, {Wang}, {Warner},
  {Willman}, {Wittman}, {Wolff}, {Wood-Vasey}, {Yoachim}, {Zhan}, \& {for the
  LSST Collaboration}}]{ive08}
{Ivezic}, Z., {Tyson}, J.~A., {Abel}, B., {et~al.} 2008, ArXiv e-prints,
  arXiv:0805.2366

\bibitem[{{Johnston} {et~al.}(2001){Johnston}, {Sackett}, \& {Bullock}}]{jsb01}
{Johnston}, K.~V., {Sackett}, P.~D., \& {Bullock}, J.~S. 2001, \apj, 557, 137

\bibitem[{{Juri{\'c}} {et~al.}(2008){Juri{\'c}}, {Ivezi{\'c}}, {Brooks},
  {Lupton}, {Schlegel}, {Finkbeiner}, {Padmanabhan}, {Bond}, {Sesar},
  {Rockosi}, {Knapp}, {Gunn}, {Sumi}, {Schneider}, {Barentine}, {Brewington},
  {Brinkmann}, {Fukugita}, {Harvanek}, {Kleinman}, {Krzesinski}, {Long},
  {Neilsen}, {Nitta}, {Snedden}, \& {York}}]{jur08}
{Juri{\'c}}, M., {Ivezi{\'c}}, {\v Z}., {Brooks}, A., {et~al.} 2008, \apj, 673,
  864

\bibitem[{{Koposov} {et~al.}(2012){Koposov}, {Belokurov}, {Evans}, {Gilmore},
  {Gieles}, {Irwin}, {Lewis}, {Niederste-Ostholt}, {Pe{\~n}arrubia}, {Smith},
  {Bizyaev}, {Malanushenko}, {Malanushenko}, {Schneider}, \& {Wyse}}]{kop12}
{Koposov}, S.~E., {Belokurov}, V., {Evans}, N.~W., {et~al.} 2012, \apj, 750, 80

\bibitem[{{K{\"u}pper} {et~al.}(2015){K{\"u}pper}, {Balbinot}, {Bonaca},
  {Johnston}, {Hogg}, {Kroupa}, \& {Santiago}}]{kue15}
{K{\"u}pper}, A.~H.~W., {Balbinot}, E., {Bonaca}, A., {et~al.} 2015, \apj, 803,
  80

\bibitem[{Law \& Majewski(2010)}]{lm10}
Law, D.~R., \& Majewski, S.~R. 2010, The Astrophysical Journal, 714, 229

\bibitem[{{Majewski} {et~al.}(2003){Majewski}, {Skrutskie}, {Weinberg}, \&
  {Ostheimer}}]{maj03}
{Majewski}, S.~R., {Skrutskie}, M.~F., {Weinberg}, M.~D., \& {Ostheimer}, J.~C.
  2003, \apj, 599, 1082

\bibitem[{{Miyazaki} {et~al.}(2012){Miyazaki}, {Komiyama}, {Nakaya}, {Kamata},
  {Doi}, {Hamana}, {Karoji}, {Furusawa}, {Kawanomoto}, {Morokuma}, {Ishizuka},
  {Nariai}, {Tanaka}, {Uraguchi}, {Utsumi}, {Obuchi}, {Okura}, {Oguri},
  {Takata}, {Tomono}, {Kurakami}, {Namikawa}, {Usuda}, {Yamanoi}, {Terai},
  {Uekiyo}, {Yamada}, {Koike}, {Aihara}, {Fujimori}, {Mineo}, {Miyatake},
  {Yasuda}, {Nishizawa}, {Saito}, {Tanaka}, {Uchida}, {Katayama}, {Wang},
  {Chen}, {Lupton}, {Loomis}, {Bickerton}, {Price}, {Gunn}, {Suzuki},
  {Miyazaki}, {Muramatsu}, {Yamamoto}, {Endo}, {Ezaki}, {Itoh}, {Miwa},
  {Yokota}, {Matsuda}, {Ebinuma}, \& {Takeshi}}]{miy12}
{Miyazaki}, S., {Komiyama}, Y., {Nakaya}, H., {et~al.} 2012, in \procspie, Vol.
  8446, Ground-based and Airborne Instrumentation for Astronomy IV, 84460Z

\bibitem[{{Newberg} {et~al.}(2009){Newberg}, {Yanny}, \& {Willett}}]{new09}
{Newberg}, H.~J., {Yanny}, B., \& {Willett}, B.~A. 2009, \apjl, 700, L61

\bibitem[{{Newberg} {et~al.}(2002){Newberg}, {Yanny}, {Rockosi}, {Grebel},
  {Rix}, {Brinkmann}, {Csabai}, {Hennessy}, {Hindsley}, {Ibata}, {Ivezi{\'c}},
  {Lamb}, {Nash}, {Odenkirchen}, {Rave}, {Schneider}, {Smith}, {Stolte}, \&
  {York}}]{new02}
{Newberg}, H.~J., {Yanny}, B., {Rockosi}, C., {et~al.} 2002, \apj, 569, 245

\bibitem[{{Newberg} {et~al.}(2003){Newberg}, {Yanny}, {Grebel}, {Hennessy},
  {Ivezi{\'c}}, {Martinez-Delgado}, {Odenkirchen}, {Rix}, {Brinkmann}, {Lamb},
  {Schneider}, \& {York}}]{new03}
{Newberg}, H.~J., {Yanny}, B., {Grebel}, E.~K., {et~al.} 2003, \apjl, 596, L191

\bibitem[{Niederste-Ostholt {et~al.}(2010)Niederste-Ostholt, Belokurov, Evans,
  \& Peñarrubia}]{no10}
Niederste-Ostholt, M., Belokurov, V., Evans, N.~W., \& Peñarrubia, J. 2010,
  The Astrophysical Journal, 712, 516

\bibitem[{{Pe{\~n}arrubia} {et~al.}(2010){Pe{\~n}arrubia}, {Belokurov},
  {Evans}, {Mart{\'{\i}}nez-Delgado}, {Gilmore}, {Irwin}, {Niederste-Ostholt},
  \& {Zucker}}]{pen10}
{Pe{\~n}arrubia}, J., {Belokurov}, V., {Evans}, N.~W., {et~al.} 2010, \mnras,
  408, L26

\bibitem[{{Sesar} {et~al.}(2017){Sesar}, {Hernitschek}, {Mitrovi{\'c}},
  {Ivezi{\'c}}, {Rix}, {Cohen}, {Bernard}, {Grebel}, {Martin}, {Schlafly},
  {Burgett}, {Draper}, {Flewelling}, {Kaiser}, {Kudritzki}, {Magnier},
  {Metcalfe}, {Tonry}, \& {Waters}}]{ses17}
{Sesar}, B., {Hernitschek}, N., {Mitrovi{\'c}}, S., {et~al.} 2017, \aj, 153,
  204

\bibitem[{{Slater} {et~al.}(2013){Slater}, {Bell}, {Schlafly}, {Juri{\'c}},
  {Martin}, {Rix}, {Bernard}, {Burgett}, {Chambers}, {Finkbeiner}, {Goldman},
  {Kaiser}, {Magnier}, {Morganson}, {Price}, \& {Tonry}}]{sla13}
{Slater}, C.~T., {Bell}, E.~F., {Schlafly}, E.~F., {et~al.} 2013, \apj, 762, 6

\bibitem[{{Vivas} {et~al.}(2001){Vivas}, {Zinn}, {Andrews}, {Bailyn}, {Baltay},
  {Coppi}, {Ellman}, {Girard}, {Rabinowitz}, {Schaefer}, {Shin}, {Snyder},
  {Sofia}, {van Altena}, {Abad}, {Bongiovanni}, {Brice{\~n}o}, {Bruzual},
  {Della Prugna}, {Herrera}, {Magris}, {Mateu}, {Pacheco}, {S{\'a}nchez},
  {S{\'a}nchez}, {Schenner}, {Stock}, {Vicente}, {Vieira}, {Ferr{\'{\i}}n},
  {Hernandez}, {Gebhard}, {Honeycutt}, {Mufson}, {Musser}, \&
  {Rengstorf}}]{viv01}
{Vivas}, A.~K., {Zinn}, R., {Andrews}, P., {et~al.} 2001, \apjl, 554, L33

\end{thebibliography}

\end{document}